\documentclass[final,1p,times,twocolumn]{elsarticle}
\def\QE{\textsc{Quantum ESPRESSO}}
\usepackage{amsmath}
\usepackage{color}
\usepackage{float}
\usepackage{amsmath}
\usepackage{array}
\usepackage{rotating}
\usepackage{algpseudocode}
\usepackage{algorithm}
\mathchardef\mhyphen="2D

\newcommand{\turbo}{\texttt{turboTDDFT}\ }
\newcommand{\gau}{Gaussian~09\ }
\def\xthinspace{\kern .1em }
\def\upl#1{\xthinspace{^{#1}}\!}

\newcommand{\SB}[1]{#1}

\begin{document}

\begin{frontmatter}

\title{turboTDDFT 2.0 -- Hybrid functionals and new algorithms
within time-dependent density-functional perturbation theory} 
\author[sissa]{Xiaochuan Ge}
\author[sissa]{Simon J. Binnie}
\author[nancy1,nancy2]{Dario Rocca}
\author[ictp]{Ralph Gebauer}
\author[sissa,epfl]{Stefano Baroni\corref{author}}

\cortext[author] {Corresponding author. \textit{e-mail address:} baroni@sissa.it}
\address[sissa]{SISSA -- Scuola Internazionale Superiore di Studi
  Avanzati, Trieste, Italy} 
\address[nancy1]{Universit\'e de Lorraine, CRM\textsuperscript{2}, UMR
  7036, Institut Jean Barriol, 54506 Vandoeuvre-l\`es-Nancy, France} 
\address[nancy2]{CNRS, CRM\textsuperscript{2}, UMR 7036, 54506
  Vandoeuvre-l\`es-Nancy, France} 
\address[ictp]{ICTP -- The Abdus Salam International Centre for
  Theoretical Physics,  Trieste, Italy} 
\address[epfl]{THEOS -- Theory and Simulation of Materials, Ecole
  Polytechnique F\'ed\'erale de Lausanne, 1015 Lausanne, Switzerland 

} 


\begin{abstract}
  We present a new release of the \turbo code featuring an
  implementation of hybrid functionals, a recently introduced
  \emph{pseudo-Hermitian} variant of the Liouville-Lanczos approach to
  time-dependent density-functional perturbation theory, and a newly
  developed Davidson-like algorithm to compute selected
  \emph{interior} eigen-values/vectors of the Liouvillian
  super-operator. Our implementation is thoroughly validated against
  benchmark calculations performed on the cyanin
  (C$_{21}$O$_{11}$H$_{21}$) molecule using the \gau and \turbo
  1.0 codes.
\end{abstract}

\end{frontmatter}
{\bf PROGRAM SUMMARY}

\begin{small}
  \noindent
  {\em Program Title:} \turbo 2.0 \\
  {\em Journal Reference:}                                      \\
  {\em Catalogue identifier:}                                   \\
  {\em Licensing provisions:} GNU General Public License V 2.0  \\
  {\em Programming language:} Fortran 95                         \\
  {\em Computer:} Any computer architecture                      \\
  {\em Operating system:} GNU/Linux, AIX, IRIX, Mac OS X, and other
  UNIX-like OS's            
  \QE, optical spectra, hybrid functionals, Lanczos recursion, Davidson
  diagonalization, pseudo-Hermitian matrix \\
  {\em Classification:} 16.2,16.6,7.7\\
  {\em External routines/libraries:} \turbo 2.0 is a tightly integrated component of the
  \QE\ distribution and requires the
  standard libraries linked by it: BLAS, LAPACK, FFTW, MPI.\\
  {\em Nature of problem:} Calculation of the optical absorption spectra of molecular
  systems.
  \\
  {\em Solution method:} Electronic excited states are addressed by linearized
  time-dependent density-functional theory within the plane-wave pseudo-potential
  method. The dynamical polarizability can be computed in terms of the resolvent of the
  Liouvillian super-operator, using a \emph{pseudo-Hermitian} variant of the Lanczos
  recursion scheme. As an alternative, individual eigenvalues of the Liouvillian can be
  computed via a newly introduced variant of the Davidson method. In both
  cases, hybrid functionals can now be used.
  \\
  {\em Restrictions:} Spin-restricted formalism. Linear-response regime. Adiabatic XC
  kernels only. Hybrid functional are only accessible using norm-conserving
  pseudo-potentials.
  \\
  {\em Unusual features:} No virtual orbitals are used, nor even calculated. Within the
  Lanczos method a single recursion gives access to the whole optical spectrum; when computing
  individual excitations usinf the Davidson method, interior eigenvalues can be easily
  targeted.
  \\
\end{small}


\section{Introduction}
\label{sec:intro}
A new approach to time-dependent (TD) density-functional theory (DFT)
\cite{Runge_1984,TDDFT_book_2012} has been recently introduced
\cite{walker_2006,rocca_turbo_2008,harnessing,our_chapter_2012},
allowing one to overcome many of the hurdles that have so far hindered
the application of TDDFT to molecular models larger than a few \SB{dozen}
atoms. This method, implemented in the \turbo package
\cite{malcolu_turbotddft_2011} of the \QE\ distribution \cite{QE-2009},
uses a Lanczos approach to evaluating the resolvent of the TDDFT
Liouvillian, requiring no explicit calculation of the unoccupied
ground state orbitals, and casting most of the computational effort
independent of frequency. As we have shown previously, this allows the
calculation of the full TDDFT absorption spectra to high precision and
at a computational cost comparable to that of a single ground state
calculation.

So far our implementation has been based around traditional LDA- or
GGA-like quasi-local exchange-correlation (XC) functionals (within the
so called \emph{adiabatic} approximation). These functionals have been
the mainstay of DFT since its inception, however they are not without
their flaws. When used in ground-state calculations, quasi-local XC
functionals are unable to capture the very essence of interaction
effects in strongly correlated materials and in weakly bound molecular
systems where intrinsically non-local correlations, responsible for
dispersion forces, play a fundamental role \cite{Weitau}. When used to
predict excited-state properties, adiabatic quasi-local XC kernels
fail to account for the long-range tail of the electron-hole (eh)
interaction, which is essential to describe charge-transfer
excitations, as well as excitons in solids and Rydberg series in
molecules \cite{Dreuw_2003,rocca_ab_2010,Ghosh_2010}.  The inability
of adiabatic XC density functionals to properly describe
charge-transfer excitations has been thoroughly discussed,
\emph{e.g.}, in Ref. \cite{Dreuw_2003} and a same analysis can be easily
generalized to excitons in insulators and Rydberg states in atoms
and molecules. What all these seemingly different phenomena have in
common is that the energy of the excited states (or their very
existence, in the case of Rydberg states and weakly bound excitons) is
determined by the long-range, Coulombic, tail of the eh
interaction. As the linearization of any local potential inevitably
results in a short-range eh interaction, it turns out that the
non-locality of the Fock exchange potential is essential to guarantee
the proper large-distance behavior of this interaction. We conclude
that in order to capture the effects of this behavior, TDDFT has to be
supplemented either with frequency-dependent XC kernels
\cite{Onida_2002} (which by definition cannot be derived from any
adiabatic functionals) or with an explicitly non-local (Fock)
component for the exchange potential, EXX. The latter route is
followed by adopting hybrid functionals \cite{Becke_1993}, in which
the exchange energy is assumed to be a linear combination of a
quasi-local (GGA-like) and a fully non-local (Fock-like)
component. The equations resulting from the linearization of TDDFT
with hybrid functionals are formally analogous to those resulting from
the linearization of the TD Hartree-Fock (HF) equation
\cite{McLachlan_1963}, which in turn resembles the Bethe-Salpeter
equation (BSE) \cite{Onida_2002} of many-body perturbation
theory. Technically, the BSE differs from linearized TDHF by the
screening of the exchange operator. The Liouville-Lanczos approach to
TDDFT has been extended so as to cope with the non-local (screened)
Fock operator and to solve the BSE in
Refs.~\cite{rocca_ab_2010,rocca_ab_2012}. Building on this work, in
this paper we introduce an implementation of the Liouville-Lanczos
approach to linearized TDDFT encompassing hybrid functionals.

As efficient as it may be, this approach has a few problems on its
own. \SB{On} the one hand, the \emph{Lanczos bi-orthogonalization
  algorithm} on which it is based is subject to
\emph{quasi-breakdowns} whenever a newly generated pair of (left and
right) vectors are almost orthogonal. On the other hand, one may be
interested in the properties of a few individual eigen-triplets,\footnote{In
  a non-Hermitian eigenvalue problem an eigen-triplet is defined as
  the set of an eigenvalue and the corresponding left and right
  eigenvectors.} rather than in the full frequency-dependent response,
or the response function may be well represented by a few
eigen-triplets only. In this case a method to compute individual
eigen-triplets with a low computational cost would be extremely
useful.

In this paper, all the above problems are addressed by introducing a
new version of the \turbo code featuring the implementation of hybrid
functionals, a recently presented \emph{pseudo-Hermitian} variant of
the Liouville-Lanczos approach to time-dependent density-functional
perturbation theory (DFPT) \cite{gruning_2011}, and a newly proposed
Davidson-like \cite{davidson_1975} algorithm to compute individual
eigen-triplets of the Liouvillian super-operator, and by making the
new code available as a component of the \QE\ distribution
\cite{QE-2009}. The paper is organized as follows: in
Sec. \ref{sec:theory} we review the general framework of the Liouville
approach to linearized TDDFT, with emphasis on hybrid functionals; in
Sec. \ref{sec:algorithms} we present the new algorithms
(\emph{pseudo-Hermitian} and Davidson's) that are being implemented in
\turbo 2.0; Sec. \ref{sec:codever} reports on the validation of the
new version of the code; Sec. \ref{sec:disc} finally contains our
conclusions.

\section{Theory}
\label{sec:theory}

\subsection{\turbo with regular (non-hybrid) density
  functionals} \label{sec:theory1} 
As outlined in previous work
\cite{our_chapter_2012,malcolu_turbotddft_2011}, the 
starting point for our TDDFT implementation is the linearized quantum
Liouville equation,
\begin{equation}
  \label{eq:lin_quantliou}
  (\omega - \mathcal{L}) \cdot \hat{\rho}^\prime(\omega) = \left[
    \hat{V}^\prime_{\rm ext}(\omega),\hat{\rho}^\circ \right],
\end{equation}
where $[\cdot,\cdot]$ indicates the commutator, $\hat{V}^\prime_{\rm
  ext}(\omega)$ is the external perturbation, and the Liouvillian
super-operator is defined as:
\begin{equation}
  \label{eq:l_pdash}
  \mathcal{L} \cdot \hat{\rho}^\prime = [\hat{H}^\circ,\hat{\rho}^\prime
  ] + [\hat{V}_{\rm  HXC}^\prime[\hat{\rho}^\prime],\hat{\rho}^\circ ].
\end{equation}
Here $\hat{H}^\circ$ and $\hat{\rho}^\circ$ are the ground-state
Kohn-Sham (KS) Hamiltonian and the corresponding ground-state density
matrix, with $\hat{V}_{\rm HXC}^\prime[\hat{\rho}^\prime]$ being the
first order perturbation of the Hartree-plus-XC potential induced by
the response density matrix. Here and in the following quantum
mechanical operators will be denoted by a caret,
``$ \hat{\phantom{A}} $''. Specifically we are interested in the case where the
external perturbation is a homogeneous electric field, allowing us to
represent the dynamical polarizability $\alpha_{ij}(\omega)$ in terms
of the dipole operator $\hat{X}_i$ and the resolvent of the
Liouvillian as
\begin{equation}
  \label{eq:polarizability}
  \alpha_{ij}(\omega) = - \left(\hat{X}_i, (\omega -
    \mathcal{L})^{-1}\cdot\left[\hat{X}_j,\hat{\rho}^\circ \right]
  \right), 
\end{equation}
where \SB{$(\cdot,\cdot)$} is a suitably defined scalar product in
operator space
\cite{rocca_turbo_2008,our_chapter_2012,malcolu_turbotddft_2011}. A
judicious choice in the representation of these (super-) operators
allows us to avoid ever having to calculate the unoccupied
ground-state orbitals. A full discussion of this can be found in Refs.
\cite{rocca_turbo_2008,our_chapter_2012,malcolu_turbotddft_2011} but
for the present purposes it is enough to note that if we define two
sets (\emph{batches}) of orbital response functions,
$\{x_v\}=\{\varphi_v^\prime({\bf r}, \omega)\}$ and
$\{y_v\}=\{\varphi_v^{\prime *}({\bf r}, -\omega)\}$, the response of
a spin-paired system can be defined as:
\begin{align}
  \label{eq:resp_density}
  \rho_\parallel^\prime({\bf r},{\bf r}^\prime,\omega) &= 
  \sum_v\bigl(x_v({\bf r})\varphi_v^{\circ *}({\bf r}^\prime) + 
    \varphi_v^{\circ}({\bf r})y_v({\bf r}^\prime)\bigr ) \\
    \rho^\prime({\bf r},{\bf r}^\prime,\omega) &= 2
    \rho_\parallel^\prime({\bf r},{\bf r}^\prime,\omega),
\end{align}
where $\rho'_\parallel$ defines the response of the ``same-spin''
density matrix, whereas $\rho'$ indicates the response summed over the
spin degrees of freedom.
Finally, a
45$^\circ$ rotation gives us the following \emph{standard batch
  representation} (SBR) for the response orbitals,
\begin{align}
  \label{eq:batch1_rot}
  q_v({\bf r}) =& \frac{1}{2}\bigl (x_v({\bf r})+y_v({\bf r}) \bigr ),\\
  \label{eq:batch2_rot}
  p_v({\bf r}) =& \frac{1}{2}\bigl (x_v({\bf r})-y_v({\bf r}) \bigr ).
\end{align}
Analogously, we can represent a generic
operator $\hat{A}$ in terms of two similar batches,
\begin{align}
  \label{eq:batch_op_pq}
  a^q_v({\bf r}) =& \frac{1}{2}\left[ \hat{Q}_c\hat{A}\varphi_v^\circ({\bf
      r}) + (\hat{Q}_c\hat{A}^\dagger\varphi_v^\circ({\bf r}))^*\right],\\
  a^p_v({\bf r}) =& \frac{1}{2}\left[ \hat{Q}_c\hat{A}\varphi_v^\circ({\bf
      r}) - (\hat{Q}_c\hat{A}^\dagger\varphi_v^\circ({\bf r}))^*\right],
\end{align}
where $\hat{Q}_c$ is the projector onto the unoccupied manifold. Only
the occupied ground-state orbitals need to be known to apply
$\hat{Q}_c$ to a set of orbitals. 
The sets of orbitals $Q=\{q_v\}$ and $P=\{p_v\}$, $(Q,P)$, and
  $A^q=\{a^q_v\}$ and $A^p=\{a^p_v\}$, $(A^q,A^p)$, are called the
  \emph{standard batch representation} (SBR) of the response density
  matrix and of the $A$ operator, respectively. This SBR has the
properties that a general real Hermitian operator has a zero $p$ or
\emph{lower} component and the commutator of said operator with the
ground-state density matrix has a zero $q$ or \emph{upper} component.
In this representation the action of the Liouvillian on the response
density matrix is given by,
\begin{equation}
  \label{eq:L_times_rho_prime}
  \mathcal{L}\cdot \hat{\rho}^\prime =
  \left(\begin{array}{cc} 0 & \mathcal{D} \\
      \mathcal{D}+ 2\mathcal{K}& 0 \end{array}\right) \left(
    \begin{array}{c} Q \\ P \end{array} \right),
\end{equation}
with the super-operators $\mathcal{D}$ and $\mathcal{K}$ defined
as,
\begin{align}
  \mathcal{D}\{ q_v({\bf r}) \} =&
  \{(\hat{H}^\circ-\epsilon_v^\circ)q_v({\bf r}) \} \label{eq:D}\\
  \mathcal{K}\{ q_v({\bf r}) \} =& \left\{\hat{Q}_c \sum_{v^\prime} 2
    \int \left( \frac{1}{|{\bf r}-{\bf r^\prime}|} +\kappa_{\rm
        XC}({\bf r},{\bf r}^\prime)\right) \varphi^\circ_v({\bf
      r})\varphi^\circ_{v^\prime}({\bf r}^\prime)q_{v^\prime}({\bf
      r}^\prime) d{\bf r^\prime}\right\}, \label{eq:K}
\end{align}
allowing Eq. \ref{eq:lin_quantliou} to be re-expressed as:
\begin{equation}
  \label{eq:sbr_quantliou}
  \left(\begin{array}{cc} \omega & -\mathcal{D} \\
      -\mathcal{D}- 2\mathcal{K}& \omega \end{array}\right) \left(
    \begin{array}{c} Q \\ P \end{array} \right)=\left(
    \begin{array}{c} 0 \\ \{ \hat{Q}_cV_{\rm ext}({\bf r})\varphi_v^\circ({\bf
        r})  \} \end{array} \right).
\end{equation}
Hartree atomic units ($\hbar=m=e=1$) are used throughout.  

We note
that, once the response density matrix is represented as in
Eq. \eqref{eq:resp_density}, by simply requiring that the $q$ and $p$
functions are orthogonal to the occupied-state manifold, the action of
the Liouvillian on it (as in Eqs. \eqref{eq:L_times_rho_prime} or
\eqref{eq:sbr_quantliou}) can be \SB{evaluated} without any reference
to unoccupied states, much in the spirit of DFPT
\cite{baroni_1987,baroni_2001}, \SB{of which the present approach to
  TDDFT is the natural extension to the dynamical regime.
  Eqs. (\ref{eq:L_times_rho_prime}-\ref{eq:K}) are all that is
  required to compute selected eigen-triplets of the Liouvillian
  super-operator using iterative techniques, such as Davidson's that
  is briefly reviewed in Sec. \ref{sec:turbo-david}, or} to evaluate
\SB{directly} the dynamical polarizability as an off-diagonal matrix
element of the resolvent of the Liouvillian (see
Eq. \eqref{eq:polarizability}), using the Lanczos bi-orthogonalization
scheme \cite{rocca_turbo_2008,Golub_1996}, \SB{briefly reviewed in
  Sec. \ref{sec:LBOA}.}

\subsection{Extending to hybrid functionals}
\label{sec:hybrids}
In hybrid functionals \cite{Becke_1993} a
fraction of the local exchange potential derived from a
(semi-) local functional is replaced by a same fraction of the
non-local Fock potential. Explicitly, the action of the ground-state
KS Hamiltonian over a test function $\bar\varphi(\mathbf{r})$
takes here the form:
\begin{equation}
  \label{eq:hybrid_h}
  \hat{H}^{\circ} \bar\varphi(\mathbf{r}) = \left( -\frac{1}{2}\nabla^2 + V_{\rm
      H}(\mathbf{r}) + V_{\rm XC}^{\mathrm{l}\alpha}({\bf r}) + V_{\rm
      ext}({\bf r} ) \right) \bar\varphi(\mathbf{r}) + \alpha \int
  V_{\rm EXX}(\mathbf{r},\mathbf{r}') \bar\varphi(\mathbf{r}') d
  \mathbf{r}', 
\end{equation}
where $V_{\rm H}$ and $V_{\rm ext}$ are the usual Hartree and external
potential potentials and $V_{\rm XC}^{\mathrm{l}\alpha}$ is a
shorthand for all the local
contributions to the XC potential. For example, in the case of
the PBE0 functional~\cite{adamo} we have $V_{\rm
  XC}^{\mathrm{l}\alpha}=V_{\rm C}^{\rm PBE}+(1-\alpha) V_{\rm
  X}^{\rm PBE}$, where $\alpha=1/4$ and $V_{\rm X}^{\rm PBE}$ and
$V_{\rm C}^{\rm PBE}$ are the PBE exchange and correlation potentials,
respectively~\cite{pbe}.  The $V_{EXX}$ kernel in Eq.~\eqref{eq:hybrid_h}
is defined as:
\begin{equation}
  \int V_{\rm EXX}(\mathbf{r},\mathbf{r}')\bar\varphi({\bf r'}) d{\bf
    r^\prime} = -\sum_{i=1}^{N} \varphi_i({\bf r}) \int \frac{\varphi_i({\bf
      r^\prime})\bar\varphi({\bf r^\prime})}{|{\bf r}-{\bf r^\prime}|} d{\bf
    r^\prime}.
  \label{eq:exx_op}
\end{equation}
In the following the local Hartree and XC contributions to the KS
potential will be globally referred to as $V^\alpha_{\rm
  HXC}(\mathbf{r})=V _{\rm H} (\mathbf{r})+V_{\rm
  XC}^{\mathrm{l}\alpha}(\mathbf{r})$, and the corresponding
contribution to the $\mathcal{K}$ super-operator (Eq. \ref{eq:K})
as $\mathcal{K}^\alpha$.

Similarly to the discussion in Sec.~\ref{sec:theory1}, we can define a
linearized quantum Liouville equation for
hybrid
functionals, resulting in the
Liouvillian super-operator:
\begin{equation}
  \label{eq:l_pdash_exx}
  \mathcal{L} \cdot \hat{\rho}^\prime = [\hat{H}^\circ,\hat{\rho}^\prime 
  ] + [\hat{V}_{\rm  HXC}^{\alpha\prime}[\hat{\rho}^\prime],\hat{\rho}^\circ ]
  + \alpha [\hat{V}_{\rm
    EXX}^\prime[\hat{\rho}^\prime],\hat{\rho}^\circ ]. 
\end{equation}
Following the lines of Ref.  \cite{rocca_ab_2010} and using a similar notation,
within the SBR of Eqs.~\ref{eq:batch2_rot}, the
Liouvillian super-operator takes the form:
\begin{equation}
  \mathcal{L}\cdot \hat{\rho}^\prime =
  \left(\begin{array}{cc} 0 & \mathcal{D}-\alpha(\mathcal{K}^{\rm
        1d}-\mathcal{K}^{\rm 2d}) \\
      \mathcal{D}+ 2\mathcal{K}^\alpha - \alpha (\mathcal{K}^{\rm 1d} +
      \mathcal{K}^{\rm 2d}) & 0 \end{array}\right) \left(
\begin{array}{c} Q \\ P \end{array} \right) ,
\label{eq:Lx_times_rho_prime}
\end{equation}
where the super-operators $\mathcal{K}^{\rm 1d}$ and $\mathcal{K}^{\rm 2d}$
are defined as:
\begin{align}
  \mathcal{K}^{1d}\{q_v(\mathbf{r})\} &=
  \left\{
    \hat{Q}_c\sum_{v^\prime} \int
    \frac{ \varphi_v^\circ({\bf r^\prime}) \varphi_{v^\prime}^\circ({\bf
        r^\prime})}{|{\bf r}-{\bf r^\prime}|} q_{v^\prime}({\bf r})d{\bf
      r^\prime}  
  \right\} \label{eq:K1d}\\
  \mathcal{K}^{2d}\{q_v(\mathbf{r})\} &=
  \left\{
    \hat{Q}_c\sum_{v^\prime} 
    \int \frac{ \varphi_v^\circ({\bf
        r^\prime}) q_{v^\prime}({\bf r^\prime})}{|{\bf r}-{\bf
        r^\prime}|} \varphi_{v^\prime}^\circ({\bf
      r})d{\bf r^\prime}
  \right\}. \label{eq:K2d}
\end{align}
Eq. \eqref{eq:Lx_times_rho_prime} has the same structure as
Eq. \eqref{eq:L_times_rho_prime}, resulting in a linear equation for
the density matrix response analogous to Eq. \eqref{eq:sbr_quantliou}.
As before this representation contains no explicit references to the
unoccupied states and can be dropped straight into our existing
Lanczos implementation for calculating generalized susceptibilities,
as well as in the new algorithms presented in
Sec. \ref{sec:algorithms}.

\subsection{Lanczos bi-orthogonalization algorithm} \label{sec:LBOA}
Once a representation has been established for the response density
matrix and for the action of the Liouvillian on it, any off-diagonal
matrix element of the resolvent of the Liouvillian, such as in
Eq. \eqref{eq:polarizability}, can be computed using the Lanczos
bi-orthogonalization algorithm (LBOA), as explained, \emph{e.g.} in
Ref. \cite{our_chapter_2012}. In a \SB{nutshell,} the LBOA amounts to the
following.

Suppose one wants to compute the \emph{susceptibility}:
\begin{equation}
  g(\omega)=\left ( u, (\omega-L)^{-1}v \right ), \label{eq:g_of_omega}
\end{equation}
where $u$ and $v$ are vectors in an $n$-dimensional linear space, and
$L$, \emph{the Liouvillian}, is an $n\times n$ matrix defined
therein. Given a pair of vectors, $u^{1}$ and $v^{1}$ normalized by
the condition $(u^{1},v^{1})=1$ (although not strictly necessary, we
assume that $u^1=v^1=v$), a pair of bi-orthogonal bases is generated
through the recursion illustrated in Algorithm \ref{alg:LBOA}.

\begin{algorithm}[h!]
  \caption{ LBOA: Lanczos bi-orthogonalization algorithm.}
  \label{alg:LBOA}
  \begin{algorithmic}
    \State $  \gamma_1v_{0}\gets 0;\quad \beta_1u_{0} \gets 0 $
    \State $ v_{1}\gets v; \quad u_{1} \gets  v $
    \State $l \gets 1$
    \Repeat
    \State $\alpha^{l}  \gets (u^{l},L v^{l}) $
    \State $ \bar v  \gets Lv_{l}-\alpha_{l}v_{l}-\gamma_{l}v_{l-1} $
    \State $ \bar u \gets L^{\top}u_{l}-\alpha_{l}u^{l}-\beta_{l}u_{l-1} $
    \State $ \beta_{l+1} \gets \sqrt{(\bar u,\bar v)} $
    \State $\gamma_{l+1} \gets \mathrm{sgn}(\bar{u},\bar{v})\times
    \beta_{l+1}$
    \State $v_{l+1} \gets \bar v/\beta_{l+1}; \quad u_{l+1} \gets
    \bar u/\gamma_{l+1} $
    \State $l \gets l+1$
    \Until {convergence or breakdown}
  \end{algorithmic}
\end{algorithm}

The \emph{oblique projection} of the Liouvillian onto the
bi-orthogonal bases, $\{u_l\}$ and $\{v_l\}$, \emph{i.e.} the matrix
$T_{ij}=(u_i,Lv_j)$, is tridiagonal. This property can be used to
compute very efficiently approximations to the susceptibility,
Eq. \eqref{eq:g_of_omega}, that can be systematically improved by
increasing the dimension of the bases, \emph{i.e.} by increasing the
number of iterations of the recursion of Algorithm \ref{alg:LBOA}
\cite{rocca_turbo_2008,harnessing,malcolu_turbotddft_2011,our_chapter_2012}.

If $(\bar{u},\bar{v}) = 0$, one says that a \emph{breakdown} has
occurred and the algorithm must be stopped. In practice, breakdowns
never occur, but \emph{quasi-breakdowns} ($(\bar{u},\bar{v}) \approx
0$) are not infrequent, resulting in spikes in the values of the
$\beta$ and $\gamma$ coefficients.  Quasi-breakdowns are sensitive to
small variations in the Liouvillian matrix, including those determined
by the use of a different arithmetics. Fortunately, the overall
algorithm results to be rather robust with respect to
them. Notwithstanding their occurrence remains an unpleasant feature
of the method.

\subsection{Numerical considerations}
\label{sec:compcons}

Looking back at Eq. (\ref{eq:Lx_times_rho_prime}) one sees that each
build of the Liouvillian super-operator requires three applications of
a non-local Fock-like operator to a batch of orbitals. Specifically,
each batch must be operated on by the $\mathcal{D}$ super-operator
(that requires the application $V_{\rm EXX}$ operator in the original
ground-state Hamiltonian on each orbital in the batch), and by the
$\mathcal{K}^{\rm 1d}$ and $\mathcal{K}^{\rm 2d}$ super-operators,
that require a similar numerical work (see Eqs. \ref{eq:K1d} and
\ref{eq:K2d}). \SB{Specializing our discussion to a plane-wave (PW)
  representation, that is adopted in the \turbo\ code presented here,
  we see that applying} a Fock-like operator to a single molecular
orbital requires two fast Fourier transforms (FFTs). So it is clear
that for any \SB{reasonably} sized system described by a reasonable number
of PWs the time taken to perform the FFTs will dominate the wall time
for the calculation, resulting in substantial aggravation of the
numerical burden, as compare with semi-local functionals.  However it
is known that the accuracy of hybrid ground-state calculations is
rather insensitive to the number of PWs used to evaluate the action of
the $V_{\rm EXX}$ operator \cite{umari_2013}. Analogously, the number
of PWs used to implement various Fock-like operators in response
calculations can be considerably reduced without an appreciable
degradation of the overall accuracy.

The way this is achieved in the \turbo code is that the
wavefunction-like objects appearing in the exchange integrals are
first interpolated onto a coarse FFT grid, with an energy cut-off that
is a fraction of the wavefunction energy cut-off used elsewhere in the
calculation. Then the density-like wavefunction products are
calculated on a real-space grid corresponding to four times the
reduced energy cut-off, to avoid any aliasing errors. The result is
then interpolated back onto the full grid and added to the appropriate
response orbitals. This whole procedure is controlled by one input
variable called {\texttt{ecutfock}} which takes as its argument the
cut-off of the reduced grid for the density-like terms. So an
{\texttt{ecutfock}} of 100 Ry would mean the wavefunction-like terms
are represented on a grid with and energy cut-off of 25 Ry and the
density-like terms on a real space grid equivalent to a 100 Ry
cut-off. We have generally found that a coarse grid with its cut-offs set to a
quarter of those of the usual grid can be used with only minimal loss
in accuracy in the final spectra. For a system dominated by the FFT
calculations this can produce a considerable overall speedup.

\subsubsection{Parallelization}

Obviously in order for the code to be useful on modern high
performance computing machines efficient parallelization is required
in order to utilize the large number of compute cores available. As
with our previous non-hybrid work we use the parallelization over
G-vectors present throughout the \QE\ distribution \cite{QE-2009}.  In
the case of EXX calculations a further level of parallelization is also
available in the ground state code. Here the total number of
processors are first divided up into a number of so-called
band-groups. Each band-group acts almost as an independent calculation
apart from the fact it only performs a certain section of the sum over
bands in Eq. (\ref{eq:exx_op}). As the calculation of this exact
exchange term dominates the total computation time this distribution
of work allows a large number of cores to be utilized efficiently.  We
have carried through this level of parallelization to our \turbo code
and each one of the exchange-like operators, Eqs. (\ref{eq:exx_op},
\ref{eq:K1d}, and \ref{eq:K2d}), are distributed in
this way. The parallelization is controlled through the command line
flag {\texttt{-nb} $N$} which can be supplied when running the
code. Here $N$ is the number of band-groups desired.

\subsection{Other considerations} Whilst the implementation of the
Lanczos recursion itself is largely unaffected by the inclusion of the
EXX potential, with only the definition of the Liouvillian being
modified, there is a secondary concern brought about the non-local
exchange operator. In order to evaluate the commutators in
Eqs. \eqref{eq:lin_quantliou} or \eqref{eq:polarizability} the dipole
operator has to be applied to the ground-state orbitals, which is an
ill-defined operation in periodic boundary conditions (PBC)
\cite{baroni_2001}.  Only the projection of the resulting orbitals
onto the empty-state manifold, however, is actually needed, and this
difficulty can be solved by the tricks usually adopted in
time-independent DFPT \cite{baroni_2001}. The latter basically amount
to transforming the action of the dipole operator into that of an
appropriate current, which is well defined in PBC and whose
implementation requires the evaluation of the commutator of the dipole
with the unperturbed KS Hamiltonian \cite{our_chapter_2012}, including
the non-local exchange potential.
For finite systems, however, this difficulty is trivially overcome by
restricting the range of the dipole operator to a limited region of
space comprising entirely the molecular electronic charge-density
distribution and not touching the boundary of the periodically
repeated simulation cell, and by evaluating its action in real space.
Another limitation of the present implementation of hybrid functionals
in \turbo 2.0 is that it is currently restricted to norm-conserving
pseudo-potentials.

\section{New algorithms}
\label{sec:algorithms}
Some of the algorithmic hurdles that hinder the way to the computation
of the spectrum of the Liouvillian super-operator are due to its
non-Hermitian character. In principle, general non-Hermitian operators
do not have real eigenvalues. However, the special form of the
Liouvillian, Eqs. \eqref{eq:L_times_rho_prime} or
\eqref{eq:Lx_times_rho_prime}, known in the literature as that of an
\emph{RPA Hamiltonian} \cite{thouless_1960}, guarantees that its
eigenvalues are real and occurring in pairs of equal magnitude and
opposite sign. Among the many demonstrations of this statement, one of
the most elegant and useful makes use of the concept of
\emph{pseudo-Hermiticity} \cite{gruning_2011,mostafazadeg_2002a}. A
matrix $M$ is said to be pseudo-Hermitian if there exists a
non-singular matrix $\eta$ such that $M=\eta^{-1}M^\dagger\eta$.

The most general form af an RPA Hamiltonian, such as
Eqs. \eqref{eq:L_times_rho_prime} or \eqref{eq:Lx_times_rho_prime},
is:
\begin{equation}
   L=\left ( 
    \begin{matrix}
      0 & B \\ A & 0
    \end{matrix}
  \right ), \label{eq:RPA}
\end{equation}
where $A$ and $B$ are Hermitian matrices of dimension, say, $n\times
n$. It is easily seen that:
\begin{equation}
  L=\sigma \bar L,
  \label{eq:RPA-fact}
\end{equation}
where
\begin{equation}
  \sigma =\left ( 
    \begin{matrix}
      0 & 1 \\ 1 & 0
    \end{matrix}
  \right )
  \quad\mathrm{and}\quad
  \bar L=\left ( 
    \begin{matrix}
      A & 0 \\ 0 & B
    \end{matrix}
  \right ).
  \label{eq:RPA-bar}
\end{equation}
The product of any two non singular Hermitian matrices, $X=YZ$
($Y=Y^\dagger$ and $Z=Z^\dagger$) is pseudo-Hemitian both with respect
to $\eta=Y^{-1}$ and with respect to $\eta=Z$. It follows that RPA
matrices of the form (\ref{eq:RPA}-\ref{eq:RPA-fact}) are
pseudo-Hermitian both with respect to $\sigma=\sigma^{-1}$ and to
$\bar L$. In Ref. \cite{mostafazadeg_2002b} it was shown that a
necessary and sufficient condition for a diagonalizable matrix to have
a real spectrum is that it is pseudo-Hemitian with respect to a
positive definite matrix. In our case, $\sigma$-pseudo-Hermiticity
would not help because $\sigma$ is not positive-definite, but $\bar
L$-pseudo-Hermiticity does, provided the latter is positive-definite,
as it is usually the case \cite{gruning_2011}. In order to prove that
the eigenvalues occur in pairs of equal magnitude and opposite sign,
let us write the eigenvalue equation:
\begin{equation}
  \left ( 
    \begin{matrix}
      0 & B \\ A & 0
    \end{matrix}
  \right )
  \left ( 
    \begin{matrix}
      Q \\ P
    \end{matrix}
  \right )
  =\omega
  \left ( 
    \begin{matrix}
      Q \\ P
    \end{matrix}
  \right ), \label{eq:RPA-eigenvalue}
\end{equation}
as
\begin{align}
  (BA)Q&=\omega^2 Q \label{eq:BAQ} \\
  (AB)P&=\omega^2 P. \label{eq:ABP}
\end{align}
Pseudo-Hermiticity of $AB$ with respect to $A^{-1}$ and $B$ (and of $BA$
with respect to $B^{-1}$ and $A$) guarantees that $\omega^2$ is real,
whereas the positive-definiteness of $A$ and $B$ implies that
$\omega^2>0$.

\subsection{Pseudo-Hermitian Lanczos algorithm}
\label{sec:pseudo-Lanczos}
Let us now assume that $\bar L$, Eq. \eqref{eq:RPA-bar}, is
positive-definite, and take it as the metric of the linear space where
the eigenvalue problem, Eq. \eqref{eq:RPA-eigenvalue}, is defined:
\begin{equation}
  \{u,v\}\doteq \left ( u,\bar L v \right ).
\end{equation}
A complete set of vectors is said to be \emph{pseudo-orthonormal} if
they are orthonormal with respect to the $\bar L$ metric,
$\{v_i,v_j\}=\delta_{ij}$. The resolution of the identity reads in
this case:
\begin{equation}
  I = \sum_i \bar L |v_i\rangle\langle v_i| =  \sum_i
  |v_i\rangle\langle v_i| \bar L. 
\end{equation}

By inserting this relation into Eq. \eqref{eq:g_of_omega}, one obtains:
\begin{equation}
  g(\omega) = \sum_i (u,v_i) \left \{ v_i,(\omega-L)^{-1}v \right
  \}. \label{eq:g_of_omega_PH} 
\end{equation}
Following Ref. \cite{gruning_2011}, the matrix elements in
Eq. \eqref{eq:g_of_omega_PH} can be obtained via a generalized
Hermitian Lanczos method where the relevant scalar products are
performed with respect to the $\bar L$ metric, as illustrated in
Algorithm \ref{alg:PHL}, and denoted as the \emph{pseudo-hermitian Lanczos
  algorithm} (PHLA).
\begin{algorithm}[h]
  \caption{ PHL: Pseudo-Hermitian Lanczos algorithm.}
  \label{alg:PHL}
  \begin{algorithmic}
    \State $  v_{0}\gets 0 $
    \State $ v_{1}\gets v $
    \State $l \gets 1$
    \Repeat
    \State $ \beta_l \gets \sqrt{(v_l,v_l)_{\bar L}} = \sqrt{(\sigma v_l,L
      v_l)} $ 
    \State $v_l \gets v_l/\beta_l $
    \State $\alpha^{l}  \gets (v_{l},L v_{l})_{\bar L}= \bigl (
      (Lv_l),\sigma(Lv_l) \bigr ) $
    \State $ v_{l+1}  \gets Lv_{l}-\alpha_{l}v_{l}-\beta_{l}v_{l-1} $
    \State $l \gets l+1$
    \Until {convergence}
  \end{algorithmic}
\end{algorithm}
The scalar products in Eq. \eqref{eq:g_of_omega_PH} can be calculated
on the fly during the Lanczos recursion, while the matrix elements of
the resolvent of the Liouvillian, $\left \{ v_i,(\omega-L)^{-1}v \right
\}$, are simply the $(i,1)$ matrix elements of the resolvent of the
tridiagonal matrix:
\begin{equation}
  \upl{m}T=
  \left(
    \begin{array}{ccccc}
      \alpha_{1} & \beta_{2} & 0 & \cdots & 0\\
      \beta_{2} & \alpha_{2} & \beta_{3} & 0 & \vdots\\
      0 & \beta_{3} & \alpha_{3} & \ddots & 0\\
      \vdots & 0 & \ddots & \ddots & \beta_{m}\\
      0 & \cdots & 0 & \beta_{m} &
      \alpha_{m}
    \end{array}
  \right),
  \label{eq:tridiagonal}
\end{equation}
$m$ being the number of iterations in Algorithm \ref{alg:PHL}. To see
this, we write the Lanczos recursion as:
\begin{equation}
  L\upl{m}V=\upl{m}V\upl{m}T+\epsilon,
  \label{eq:Lanczos_factorization}
\end{equation}
where $\upl{m}V=[v_1,v_2,\cdots v_m]$ is an $n\times m$ matrix whose
columns are the Lanczos iterates, $\upl{m}T$ is the matrix in
Eq. \eqref{eq:tridiagonal}, and $\epsilon$ a remainder that can be
neglected as the number of iterations is large enough. Let us
define $\upl{m}U=[\bar Lv_1,\bar Lv_2, \cdots \bar
Lv_m]$. {Pseudo-orthonormality reads: $\upl{m}U^\dagger
  \upl{m}V=\upl{m}I$, where $\upl{m}I$ is the $m\times m$ identity
  matrix.} Let us subtract $\omega \upl{m}V$ from both sides of
Eq. \eqref{eq:Lanczos_factorization} and multiply the resulting
equation by $\upl{m}U^\dagger (L-\omega)^{-1}$ on the left and by
$(\upl{m}T-\omega)^{-1}$ on the right. The final result is: $
\upl{m}U^\dagger (\omega-L)^{-1} \upl{m}V = (\omega-\upl{m}T)^{-1} $,
whose $(i,1)$ matrix element, taking into account that $v=v_1$, is:
$  \left \{ v_i,(\omega-L)^{-1}v \right \}= 
\left [ (\omega-\upl{m}T)^{-1} \right ]_{i1}$.
Considering that one column of the inverse of a tridiagonal
matrix can be inexpensively calculated by solving a tridiagonal
linear system, as it is the case in the ordinary Hermitian or
non-Hermitian Lanczos methods, we have all the ingredients for an
accurate and efficient calculation of the generalized
susceptibility, Eq. \eqref{eq:g_of_omega}. The two main features
of the PHLA, that make it stand out
over the standard LBOA, are that it requires half as many matrix
builds and that, not relying on left and right vectors that can
occasionally be almost orthogonal to each other, it is not subject
to quasi-breakdowns.

\subsection{Turbo-Davidson diagonalization}
\label{sec:turbo-david}
There are a number of problems where one is interested in the
computation of individual eigen-triplets, rather than in a specific
dynamical response function over a wide energy range. These include
the computation of excited-state properties, such as \emph{e.g.}
forces in higher Born-Oppenheimer surfaces \cite{hutter_2003}, or such
small systems that their dynamical response functions can be
satisfactorily represented by a few eigen-triplets. The computation of
selected eigenvalues of large matrices can be conveniently achieved by
subspace iterative methods \cite{tretiak_2009}, of which the classical
Davidson algorithm \cite{davidson_1975} is one of the most successful
examples.  This algorithm was first formulated to solve large
Hermitian eigenvalue problems, such as occurring, \emph{e.g.}, in the
configuration-interaction method. It was then largely employed in
various self-consistent field codes, such as in \QE \cite{QE-2009}
among several others, or in the calculation of dynamical response
functions in the Tamm-Dancoff Approximation \cite{hirata_1999}, which
results in a Hermitian eigenvalue problem. The algorithm was then
later extended so as to encompass the RPA matrices occurring in
general response calculations \cite{rettrup_1982}, and further
generalized in a so-called \emph{representation independent}
formulation \cite{tretiak_2009} that is similar in spirit to our own
\emph{DFPT representation} \cite{rocca_turbo_2008}.

In order to illustrate the Davidson algorithm, we begin with the
computation of the lowest-lying eigen-pair of a Hermitian matrix, $H$.
Starting from a trial eigenvector, $v_1$, which we assume to be
normalized, $\left (v_1,v_1 \right )=1$, and the corresponding
estimate for the eigenvalue, $\epsilon_1=\left ( v_1, H v_1 \right )$,
one iteratively builds a reduced basis set according to the following
procedure. One first defines the residual of the current estimate of
the eigenvector, $r=\left (H-\epsilon_1\right )v_1$, and then appends
it to the current basis set upon orthogonalization to all its elements
and normalization. In practice, before the orthogonalization is
performed, the residual is \emph{preconditioned}, \emph{i.e.}
multiplied by a suitably defined matrix, $G$, chosen in such a way as
to speed up convergence. For diagonally dominant matrices, one often
sets $G=\left (\mathrm{diag}(H)- \epsilon_1 \right )^{-1}$. Once the
basis has been thus enlarged, $H$ is diagonalized in the subspace
spanned by it (the \emph{iterative subspace}), the lowest-lying
eigenvector is elected to be the new trial vector, and the procedure
repeated until convergence is achieved (typically, $\parallel\!
r\! \parallel$ smaller than some threshold), or the dimension of the
reduced basis set exceeds some preassigned value, and is then reset.

In the following we present an extension of this algorithm designed to
find several eigen-triplets of a Liouvillian eigenvalue problem,
Eq. \eqref{eq:RPA-eigenvalue}.  matrix. Our algorithm is a variation
of the one described in Sec. IIIB of Ref. \cite{tretiak_2009},
featuring a sorting phase, that allows to target \emph{interior}
eigenvalues (\emph{i.e.}  those that are closest in magnitude to a
preassigned reference excitation energy, $\omega_{ref}$) and a
preconditioning step that significantly increases its performance.

Let $L$ be an $n\times n$ RPA matrix of the form of Eq. \eqref{eq:RPA}
and let us suppose that one is interested in finding its $k$
eigenvalues whose magnitude is closest to some reference value,
$\omega_{ref} >0$. At variance with the LBOA of
Sec. \ref{sec:pseudo-Lanczos} and with a different choice that could
be made in the present case as well, the \emph{iterative subspace} is
defined here in terms of an orthonormal basis, rather than of a system
of bi-orthogonal bases. Let $\{w_i\}$ be such an orthonormal set of
$m$ $n$-dimensional arrays and $W=[w_1,w_2, \cdots w_m]$ the $n\times
m$ matrix whose columns are basis vectors. Orthonormality reads:
$\upl{m}W^\dagger \upl{m}W={^mI}$, where ${^mI}$ is the $m\times m$
identity matrix.  The $Q$ and $P$ components of the right eigenvector
of Eq. \eqref{eq:RPA-eigenvalue} are obtained as the right and left
eigenvectors of the matrix $C=BA$ (see Eq. \eqref{eq:BAQ}),
respectively.  Following the arguments given at the begining of this
Sec. \ref{sec:algorithms}, in order to guarantee that the reduced
eigenvalue problem has real eigenvalues, it is convenient to enforce
that the reduced matrix is pseudo-Hermitian. This is easily achieved
by projecting, instead of the product matrix $C=BA$, its factors, $A$
and $B$ \cite{define-residue}. We define therefore the reduced matrix
to be diagonalized as $\upl{m}C=\upl{m}B \upl{m}A$, where $\upl{m}A =
\upl{m}W^\dagger A \upl{m}W$ and $\upl{m}B = \upl{m}W^\dagger B
\upl{m}W$. Being interested in finding multiple interior
eigen-triplets simultaneously, the orthogonal basis sets is enlarged
by appending to it the left and right residuals corresponding to the
eigenvalues whose square roots are closest to a target reference
eigenvalue, $\omega_{ref}$. The residuals themselves are calculated
and preconditioned from Eq. \eqref{eq:RPA-eigenvalue}, rather than
from Eqs. (\ref{eq:BAQ}-\ref{eq:ABP}). The action of the
preconditioner on a batch of orbitals, $Q=\{q_v\}$, is defined as:
\begin{equation}
GQ=\{ \hat Q_c\hat g(\epsilon_v) q_v \},
\label{eq:precondition}
\end{equation}
where $ \hat g(\epsilon_v) = \left ( \mathrm{diag}(\hat H_0-\epsilon_v
  -\omega_{ref}) \right )^{-1}$, and $\hat Q_c$ is the projector over
the empty-state manifold, to satisfy the constraint that batch
orbitals be orthogonal to all the occupied states. In practice, the
unperturbed Hamiltonian $\hat H_0$ can be substituted with the
kinetic-energy operator, without any appreciable degradation of the
performances.

\begin{algorithm}[h!]
  \caption{turboDavidson: Davidson-like algorithm for multiple
    interior eigenvalues of the Liouvillian.}
  \label{alg:turbo-davidson}
\begin{algorithmic}[1]
  \State Initialization: $m\gets 2k$; $\upl{m}W\gets [w_1,w_2,\cdots
  w_m]$; 
  \Repeat 
  \State Orthonormalization: $\upl{m}W^\dagger \upl{m}W=\upl{m}I$;
  \State $\upl{m}A\gets \upl{m}W^\dagger A \upl{m}W$; $\upl{m}B \gets
  \upl{m}W^\dagger B \upl{m}W$; $\upl{m}C \gets \upl{m}B 
  \upl{m}A$;
  \State Solve: $\upl{m}C^\dagger \bar p_i = \bar \omega_i^2
  \bar p_i$ and $\upl{m}C \bar q_i = \bar \omega_i^2 \bar q_i$;
  \State Sort $\{\bar p_i,\bar q_i, \bar\omega^2_i\}$ in ascending order of 
  $|\bar\omega_i-\omega_{ref}|$;
  \For {$i=1$ to $k$} 
  \If { ($\{p_i,q_i,\omega^2_i\}$ is not converged)}
  \State Update the eigenvectors of $C$: $q_i \gets \upl{m}W \bar q_i$; $p_i
  \gets \upl{m}W \bar p_i$;
  \State Compute left and right residuals: $r_i\gets B p_i - (p_i,Bp_i)q_i$;
  $l_i\gets A q_i - (q_i,Aq_i)p_i$;
  \If {($|r_i|<\epsilon$ and $|l_i|<\epsilon$)} 
  \State $p_i$ and $q_i$ are converged; 
  \Else
  \If {($|r_i|>\epsilon$)} 
  \State $m \gets m+1$; $w_{m} \gets G r_i$;
  \EndIf
  \If {($|l_i|>\epsilon$)}  
  \State $m \gets m+1$; $w_{m} \gets G l_i$;
  \EndIf
  \EndIf
  \EndIf
  \EndFor
  \Until {(All the first $k$ eigen-triplets are converged)}
\end{algorithmic}
\end{algorithm}

Algorithm box \ref{alg:turbo-davidson} summarizes the Davidson
algorithm for the pseudo-Hermitian problem. To find $k$
eigen-triplets, we usually start from $2k$ trial vectors (line 1).  At
each iteration basis vectors are orthonormalized (line 3). The reduced
matrix $\upl{m}C$ is first generated (line 4) and then diagonalized
(line 5). In order to target interior eigenvalues, the resulting
eigen-triplets are sorted in order of ascending $|\bar
\omega_i-\omega_{ref}|$ (line 6). For each one of the $k$
eigen-triplets of reduced matrix $\upl{m}C$, one obtains an estimate
of the corresponding eigenvector in the full linear manifold (line 8)
and the residuals are estimated using Eq. \eqref{eq:RPA-eigenvalue}
(line 9); convergence is finally tested, and the reduced basis set
updated upon preconditioning of the residuals (see
Eq. \ref{eq:precondition}), when appropriate (lines 15 and 18).

\section{Code validation}
\label{sec:codever}

\begin{figure}[h!]
  \begin{center}
    \includegraphics[width=0.47\textwidth]{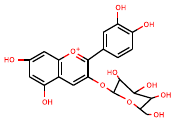}
  \end{center} 
  \caption{Chemical structure of the cyanin molecule
    (C$_{21}$H$_{21}$O$_{11}$) used to validate the new features and
    algorithms implemented in \turbo 2.0 } 
  \label{fig:cyanin}
\end{figure}

The new features and algorithms implemented in \turbo 2.0 have been
validated with the \textit{cyanin} molecule
(C$_{21}$H$_{21}$O$_{11}$: see Fig. \ref{fig:cyanin})
as a test bed, using a super-cell of $20\times20\times12$
${\AA^3}$. Standard DFT calculations were performed using the PBE
functional \cite{PBE}, ultra-soft pseudo-potentials \cite{USpseudo} from the \QE\
public repository \cite{QE_pseudo}, and a 25/200 Ry PW basis
set.\footnote{The notation ``X/Y Ry PW'' indicates a kinetic energy
  cutoff of X and Y Ry has been adopted for expanding molecular
  orbitals and the charge distributions, respectively.}  Hybrid-functional
calculations were performed with the PBE0 functional \cite{adamo},
using norm-conserving PBE pseudopotentials, and a PW cutoff of 40/160 Ry
\cite{QE_pseudo}. \SB{All the spectra were computed at the
PBE-optimized geometry\footnote{\SB{Atomic
    coordinates are available from the authors upon request.}}
 in a $20\times20\times12$
${\AA}$ tetragonal super-cell and
broadened by evaluating them at
complex frequencies $\omega+i\eta$, with $\eta=0.005$ Ry.}

\subsection{turboDavidson}

\begin{figure}[h!]
  \begin{center}
    \includegraphics[width=0.48\textwidth]{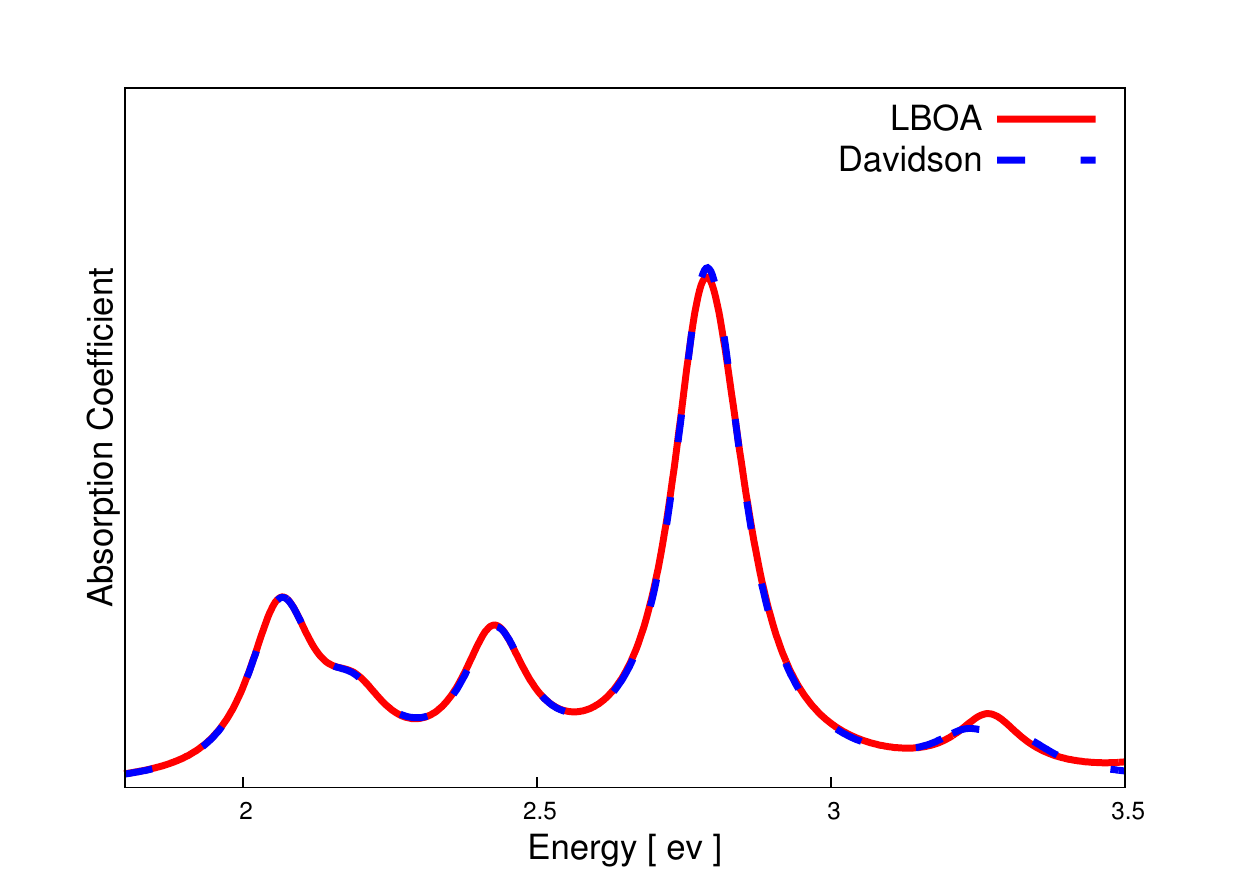}
    \includegraphics[width=0.483\textwidth]{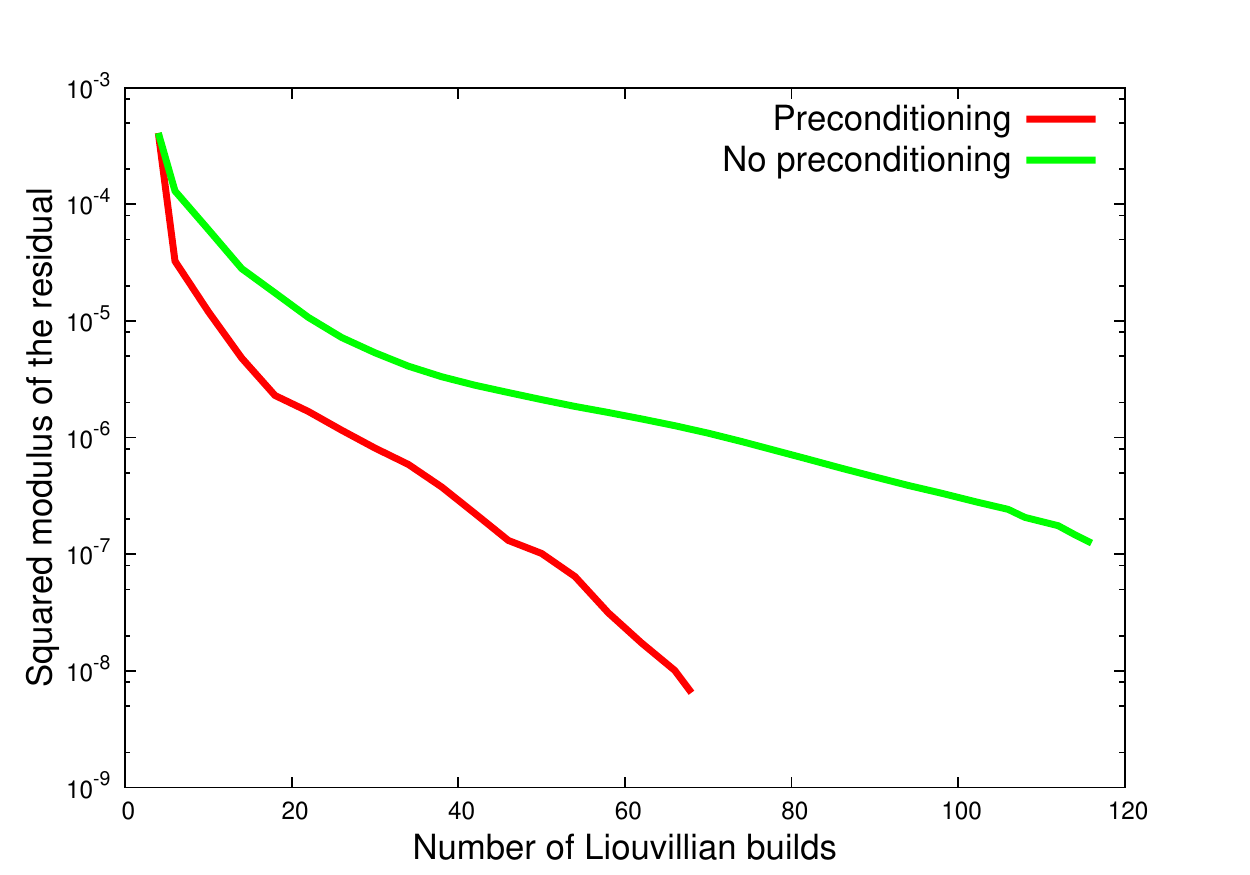}
  \end{center} 
  \caption{Left panel: Comparison of the spectra of cyanin calculated
    by the LBOA and our \emph{turboDavidson} algorithm. Right panel:
    squared norm of the residual vector for the lowest excitation
    energy of cyanin versus the number of Liouvillian builds.}
  \label{fig:davidson}
\end{figure}

In the left panel of Fig. \ref{fig:davidson} we compare the spectra
computed for cyanin in the visible range, using the LBOA and the newly
implemented turboDavidson algorithm. The LBOA calculation was
performed using 3000 Lanczos iterations per polarization, while 15
Liouvillian eigentriplets were computed with turboDavidson. 9
iterations were sufficient in this case to abate the estimated error
in the eigenvalues below $10^{-4}$ \SB{Ry}. The total number of
Liouvillian builds is twice the number of iterations in LBOA, while it
is equal to the number of (right or left) unconverged eigenvectors per
iteration in turboDavidson (less than 250 in total in present
case). This substantial saving in computer time (at the cost of a
substantial increase in the memory needed by Davidson algorithms) is
typical of all those cases where one is interested in the optical
spectrum over a restricted energy range, with only a few discrete
transitions occurring therein.

The present Davidson algorithm is a variant of the one introduced in
Ref. \cite{tretiak_2009}, featuring a newly introduced preconditioning
(see Eq. \eqref{eq:precondition}) that dramatically improves upon the
performances of the original algorithm. This improvement is
illustrated in the right panel of Fig.\ref{fig:davidson} that reports
the magnitude of the squared modulus of the residual (line 10 in
algorithm box \ref{alg:turbo-davidson}) as a function of the number of
Liouvillian builds, with and without preconditioning.

\subsection{Pseudo-Hermitian Lanczos}
\begin{figure}[h!]
  \begin{center}
    \includegraphics[width=0.48\textwidth]{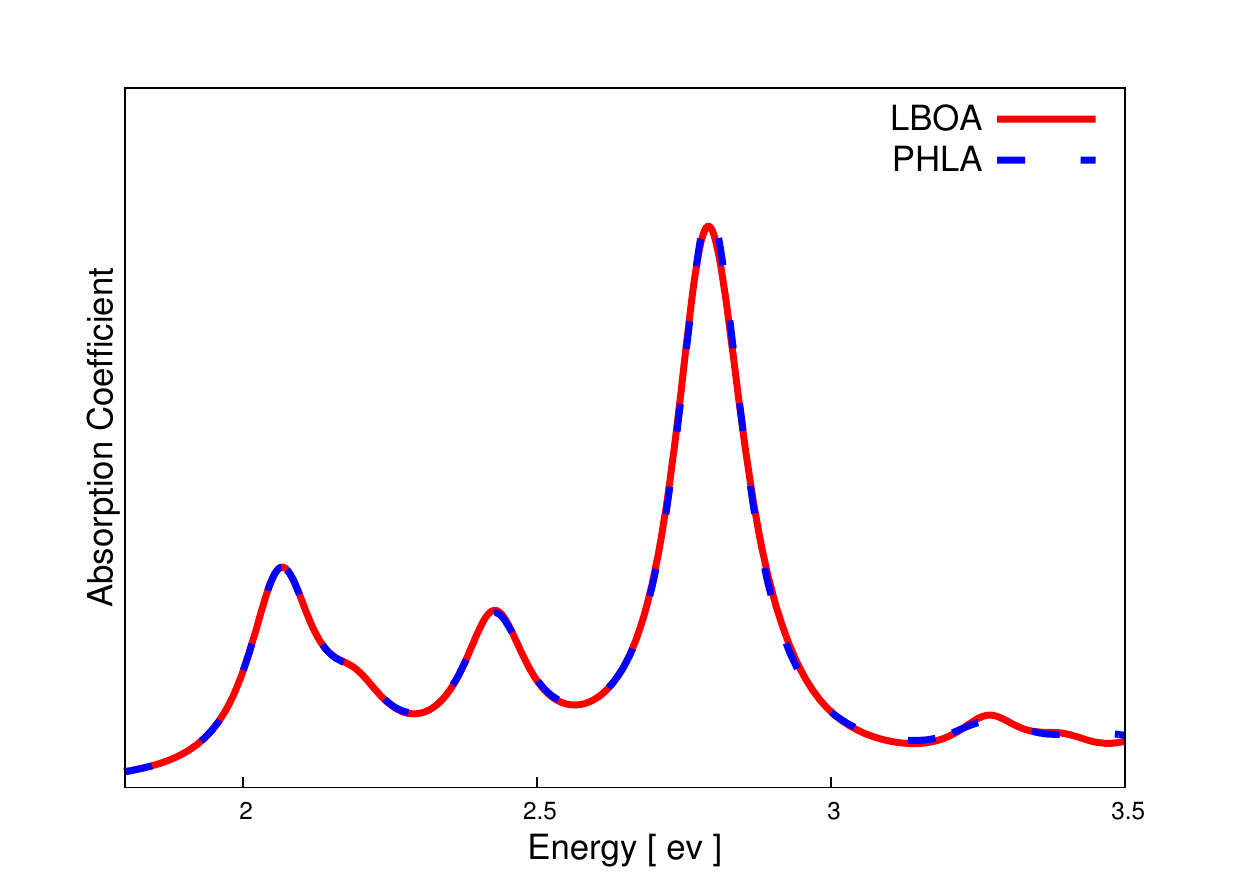}
    \includegraphics[width=0.48\textwidth]{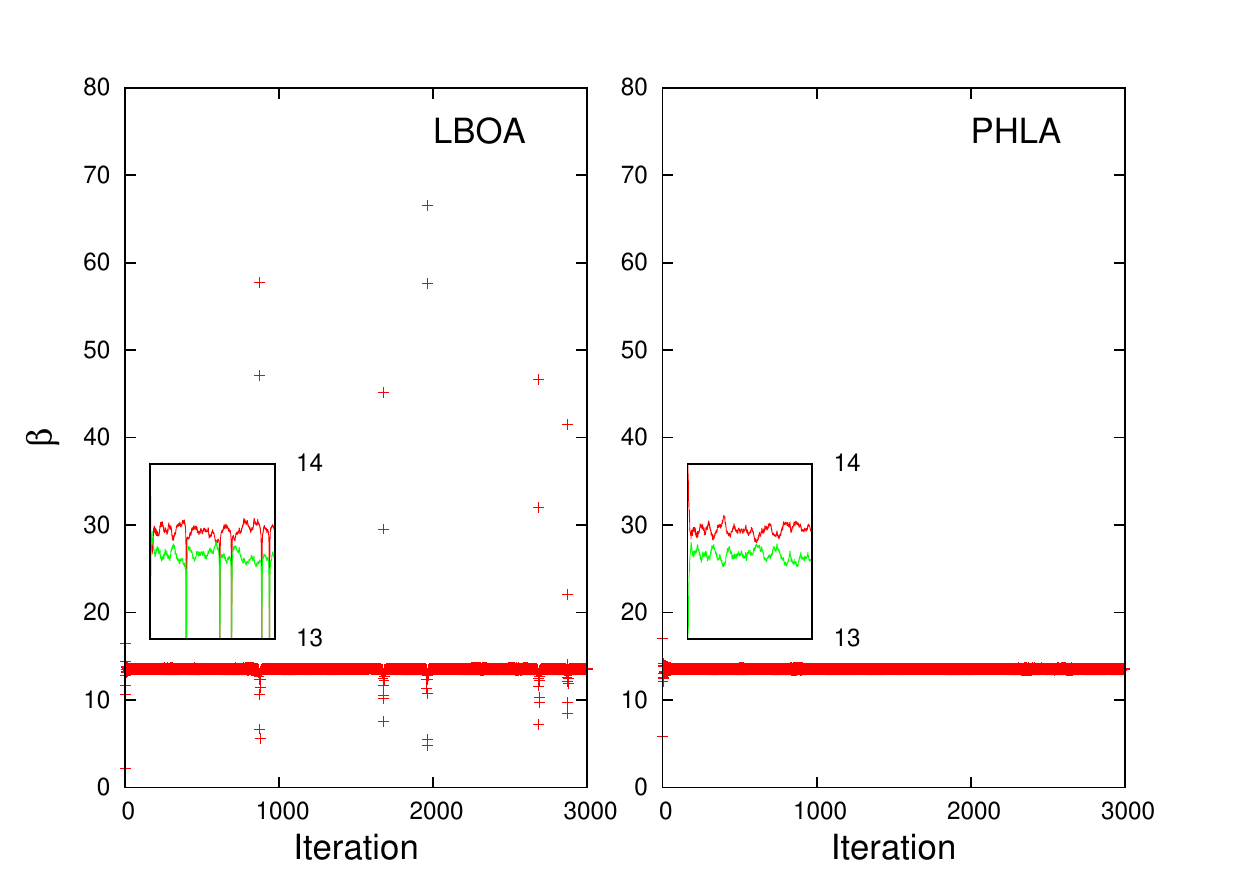}
  \end{center} 
  \caption{Left panel: Comparison of the spectra of cyanin calculated
    by the LBOA and the PHLA. Right panels: beta coefficients generated by
    the LBOA (\textit{left}) and PHLA (\textit{right}) algorithms. In
    the inset, the same data are shown on a magnified scale and with
    different colors for odd (red) and even (green) coefficients. }
  \label{fig:phla}
\end{figure}

In the left panel of Fig. \ref{fig:phla} we compare the spectra
computed for cyanin using two different flavors of the Lanczos
algorithm: the LBOA and the newly introduced PHLA. We see that the
substantial saving in the number of Liouvillian builds realized by the
PHLA (a factor of two per iteration, with respect to \SB{other} LBOA) does
not compromise on the accuracy, if not for a marginal increase in the
number of iterations needed to achieve a target accuracy (a few
percent in our experience).

As it was discussed in Sec. \ref{sec:LBOA}, the LBOA is subject to
quasi-breakdowns occurring whenever a newly generated pair of (left
and right) vectors are almost orthogonal. When a quasi-breakdown
occurs, the newly generated $\beta$ coefficient (see Algorithm box
\ref{alg:LBOA}) suddenly blows up, as illustrated in the middle panel
of Fig. \ref{fig:phla}. As it is typical of numerical instabilities in
Lanczos-like algorithms, the occurrence of quasi-breakdowns depends on
a number numerical details of the calculation (such as size of the
super-cell, the kinetic-energy cutoff, or even the computer
arithmetic). Fortunately, to the best of our knowledge and experience,
the quality of the computed spectrum is rather insensitive to these
numerical details and to the resulting instabilities. This being said,
it is relieving that such a simple fix as the the PHLA, which results
in a substantial reduction of computer time, also completely
eliminates these instabilities, as it is demonstrated in the right
panel of Fig. \ref{fig:phla}.

\subsection{Hybrid functionals}

\begin{figure}[t!]
  \begin{center}
    \hbox to \hsize{\hfill
      \includegraphics[width=0.48\textwidth,clip]{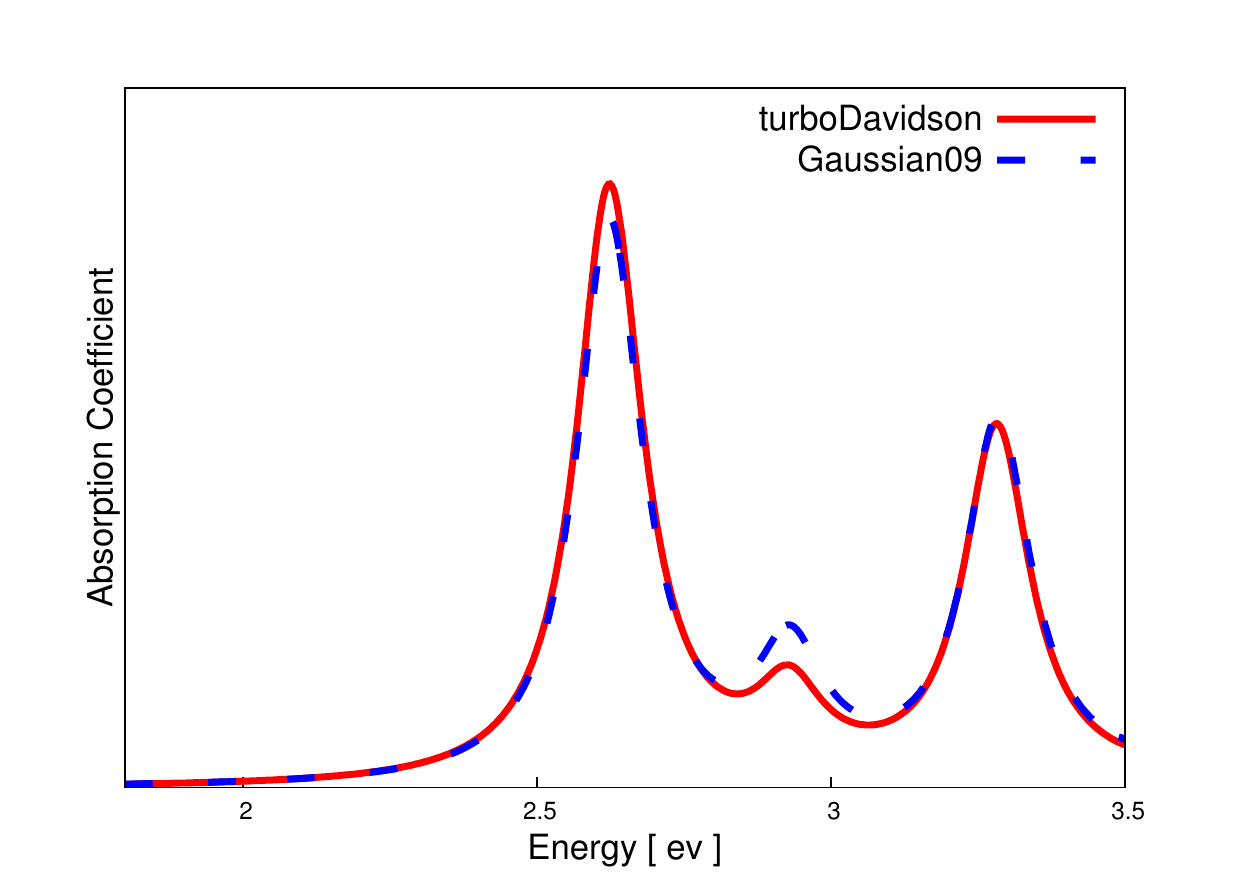}
      \hfill
      \includegraphics[width=0.48\textwidth,clip]{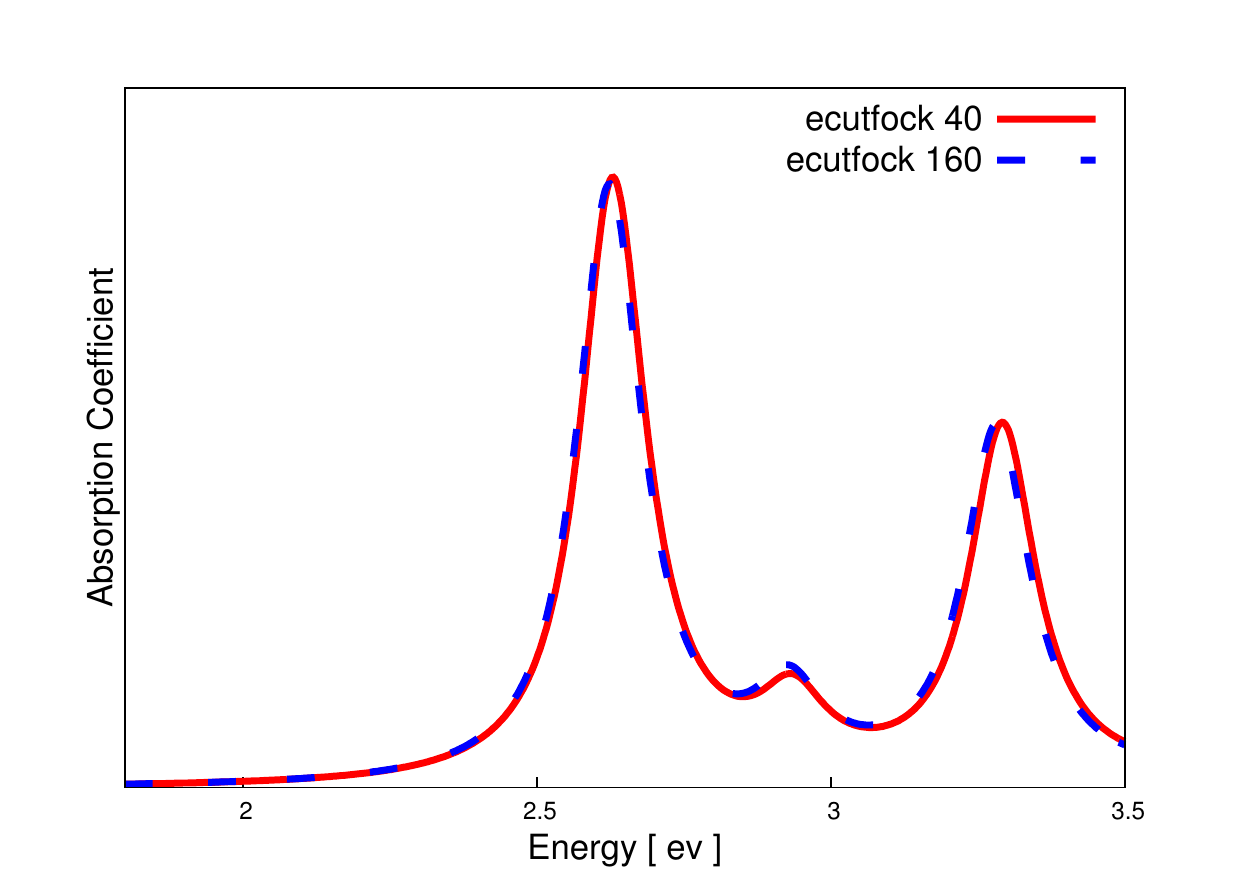}
      \hfill
    }
    \caption{Left panel: Absorption spectra of cyanin, calculate by
      turboDavidson and Gaussian09, respectively. Right panel:
      comparison of the spectra calculated with the \turbo code using
      several different values of {\texttt{ecutfock}}.}
    \label{fig:hybrid}
  \end{center}
\end{figure}

In Fig. \ref{fig:hybrid} we compare the optical spectra calculated
with the PBE0 functional using a PW basis set and the present
implementation of the Davidson algorithm with those obtained with
a Gaussian basis set and the \gau code. The two spectra match
almost perfectly, the residual discrepancies being likely due to the
slight inconsistency introduced by using a PBE pseudo-potential
\cite{QE_pseudo} with a PBE0 functional.

As the action of the $V_{EXX}$ operator requires the evaluation of the
convolution of products of pairs of one-electron wave-functions with
the Coulomb kernel (see Eq. \eqref{eq:exx_op}), one could expect that
these operations require a PW cut-off four times as large as the
wavefunction cut-off. The discussion in Sec. \ref{sec:compcons},
however, suggests that the overall accuracy of a response calculation
may be rather insensitive to the value of the PW cutoff employed for
evaluating the $V_{EXX}$ operator, \texttt{ecutfock}. In the right
panel of Fig. \ref{fig:hybrid} we display the spectra obtained using a
value for \texttt{ecutfock} ranging from the value used for the
Kohn-Sham orbitals to four times as much, and demonstrates that
setting $\texttt{ecutfock}=\texttt{ecutwfc}$ does not result in any
significant loss of accuracy, and this value is therefore assumed as
the default one.

\section{Conclusions}
\label{sec:disc}

In this paper we have introduced three substantial improvements and
generalizations to the algorithms implemented in the \turbo code for
computing excited-state properties within linearized TDDFT. First and
foremost, \turbo 2.0 allows to use hybrid functionals, which
substantially improve \SB{upon} the accuracy attainable by semilocal
PBE-like functionals. Second, it is now possible to compute individual
excitation energies and amplitudes using a variant of the Davidson
algoritm. Finally, our Liouville-Lanczos approach to the computation
of optical spectra has been substantially improved by adopting a
recently proposed pseudo-Hermitian variant to it. Work is in progress
to remove the present limitation of hybrid functionals to
norm-conserving pseudo-potentials, to unify the approaches to hybrid
functionals and to the Bethe-Salpeter equation, and to further improve
the overall numerical efficiency of the algorithms implemented in the
\turbo code.

\newpage
\appendix
\section{Input Variables}
\begin{table}[H]
    \begin{tabular}{|>{\centering}m{0.6cm}|c}
      \cline{1-1} 
      Card  & \begin{tabular}{|>{\centering}m{3.45cm}|>
          {\centering}m{1.5cm}|>{\centering}p{9.0cm}|} 
        \hline 
        Variable name  & Default Value  & Description\tabularnewline
      \end{tabular}\tabularnewline
      \cline{1-1} 
      \begin{sideways}
        \textbf{lr\_input}%
      \end{sideways}
      & \begin{tabular}{|>{\raggedright}m{3.45cm}|>
          {\centering}m{1.5cm}|m{9.0cm}|} \hline 
        \texttt{prefix} & 'pwscf' & {\footnotesize 
          The files generated by the ground state \texttt{pw.x} run should have
          this same prefix.}\tabularnewline \hline \texttt{outdir} & './' & {\footnotesize
          Working directory. On start, it should contain the files
          generated by a ground state \texttt{pw.x} run.}\tabularnewline \hline
        \texttt{wfcdir} & unset & {\footnotesize 
          Directory for scratch data.}\tabularnewline \hline \texttt{restart} & \emph{.false.}  &
        {\footnotesize When set to \emph{.true.}, \texttt{turbo\_lanczos.x} will attempt
          to restart from a previous interrupted calculation.  (see \texttt{restart\_step}
          variable).}\tabularnewline \hline \texttt{restart\_step} & \texttt{itermax} &
        {\footnotesize The code writes restart files every \texttt{restart\_step}
          iterations. Restart files are automatically written at the end of
          \texttt{itermax } Lanczos steps.} \tabularnewline \hline \texttt{lr\_verbosity}
        & 1 & {\footnotesize 
        Verbosity level}\tabularnewline \hline
      \end{tabular}\tabularnewline
      \cline{1-1} 
\begin{sideways}
  \textbf{lr\_control}  
\end{sideways}
& \begin{tabular}{|>{\raggedright}m{3.45cm}|>{\centering}m{1.5cm}|m{9.0cm}|}
  \hline \texttt{itermax } & 500 & {\footnotesize Number of iterations
    to be performed.}\tabularnewline \hline \texttt{ipol } & 1 &
  {\footnotesize 
    Component of the polarizability tensor to be computed:
    $1 \rightarrow (xx)$, $2\rightarrow (yy)$, 
  $3 \rightarrow (zz)$. When set to 4 the full polarizability
    tensor is
    computed.}\tabularnewline \hline \texttt{nipol } & {\footnotesize
    1 if ipol $<$ 4; 3 if ipol=4} & {\footnotesize Determines the
    number of zeta coefficients to be calculated for a given
    polarization direction.}  \tabularnewline \hline \texttt{ltammd }
  & \emph{.false.}  & {\footnotesize When set to \emph{.true.} the
    Tamm-Dancoff approximation is used.}\tabularnewline \hline \texttt{no\_hxc } &
  \emph{.false.}  & {\footnotesize When set to \emph{.true.} the
    response of the Hartree and XC potentials are
    ignored, resulting in an independent-electron
    approximation.}
      \tabularnewline \hline 
      {{\texttt{d0psi\_rs}}} & {{\footnotesize .false.}} &
      {{\footnotesize When set to .true., the matrix elements of the dipole
          operator are calculated in real space. Can be used only for molecules 
          or clusters, NOT for periodic systems. Molecule must be placed in the 
          centre of the simulation cell. It is recommended to set this value
          only for hybrid-functional calculations. }} 
    \tabularnewline \hline {\bf \texttt{ecutfock }} &
  $4\times \texttt{ecutwfc}$ & {\footnotesize PW
    cut-off to compute EXX-like terms.}
    \tabularnewline \hline \texttt{charge\_response } & 0 &
  {\footnotesize When set to 1, the code computes and writes the response
    charge density. Setting \texttt{charge\_response }
    to 1 makes the presence of the card \textbf{lr\_post} mandatory.}
  \tabularnewline \hline \texttt{pseudo\_hermitian} & .true. &
  \footnotesize{Setting this parameter to .false. one can use the LBOA
    as in \turbo 1.0.}  
\end{tabular}\tabularnewline
\cline{1-1} 
    \begin{sideways}
      \textbf{lr\_post}%
    \end{sideways}
    & \begin{tabular}{|>{\raggedright}m{3.45cm}|>{\centering}m{1.5cm}|m{9.0cm}|}
      \hline \texttt{omeg } & 0.0 & {\footnotesize The response of the
        charge density is calculated for this transition energy (in
        Rydberg units)}\tabularnewline \hline \texttt{epsil } & 0.0 &
      {\footnotesize The broadening/damping term (in Rydberg
        units).}\tabularnewline \hline \texttt{beta\_gamma\_z\_prefix}
      & 'pwscf' & {\footnotesize The prefix of the file where the beta
        gamma zeta coefficients from the first calculation can be set
        manually using this parameter. The file
        \texttt{outdir/beta\_gamma\_z\_prefix.beta\_gamma\_z.x} (where
        \texttt{x}=1-3) must exist.  }\tabularnewline \hline
      \texttt{w\_T\_npol } & 1 & {\footnotesize Number of polarization
        directions considered in the previous calculation. It must be
        set to \texttt{3} if in the previous calculation
        \texttt{ipol=4}, it must be set to \texttt{1}
        otherwise.}\tabularnewline \hline \texttt{plot\_type } & 1 &
      {\footnotesize An integer variable that determines the format of
        the file containing the charge density response.  1: A file
        containing the x y z grid coordinates and the corresponding
        value of the density is produced 2: The density response is
        written in Xcrysden format 3: The density response is written
        in the gaussian cube format} \tabularnewline \hline
    \end{tabular}\tabularnewline
    \cline{1-1}
\end{tabular}
\caption{Input variables for \texttt{turbo\_lanczos.x} }
\label{TableINPtddfpt.x}
\end{table}


\begin{table}[H]
    \begin{tabular}{|>{\centering}m{0.6cm}|c}
      \cline{1-1} 
      Card  & \begin{tabular}{|>{\centering}m{3.45cm}|>{\centering}m{1.5cm}|>{\centering}p{9.0cm}|}
        \hline 
        Variable name  & Default Value  & Description\tabularnewline
      \end{tabular}\tabularnewline
      \cline{1-1} 
      \begin{sideways}
        \textbf{lr\_input}%
      \end{sideways} & \begin{tabular}{|>{\raggedright}m{3.45cm}|>{\centering}m{1.5cm}|m{9.0cm}|}
        \hline 
        \texttt{prefix}  & 'pwscf'  & {\footnotesize Sets the prefix for generated and read files. The files generated by
          the ground state \texttt{pw.x} run should have this same prefix.}\tabularnewline
        \hline 
        \texttt{outdir}  & './'  & {\footnotesize The directory that contains the run critical files, which include the files 
        generated by ground state \texttt{pw.x} run.}.\tabularnewline
        \hline  
        \texttt{wfcdir}  & unset  & {\footnotesize The directory where the scratch files will be written and read. Restart
          related files are always written to outdir.}\tabularnewline
        \hline 
      \end{tabular}\tabularnewline
      \cline{1-1} 
\begin{sideways}
  \textbf{lr\_dav}  
\end{sideways} & \begin{tabular}{|>{\raggedright}m{3.45cm}|>{\centering}m{1.5cm}|m{9.0cm}|}
  \hline \texttt{num\_eign} & 1 & {\footnotesize Number of eigenstates to be calculated.}
  \tabularnewline \hline 
 \texttt{num\_init} & 2 & {\footnotesize Number of trial vectors.} \tabularnewline \hline 
 \texttt{if\_random\_init} & .false. & {\footnotesize When set to $.true.$ trial vectors
   are chosen randomly, otherwise they are guessed from ground-state calculation. If
   $N_vN_c < num\_init$, this term is forced to be $.true.$.}  \tabularnewline \hline  
 \texttt{num\_basis\_max } & 20 &
 {\footnotesize Maximum number of basis vectors allowed in the
   subspace. When this number 
   is reached, a discharging routine is called, sorting out $2\times num\_eign$ best
   vectors in the subspace and dumping the others.} \tabularnewline \hline 
 \texttt{reference } & 0.0 &
 {\footnotesize $\omega_{ref}$ (\SB{Ry}) is used to favor these eigenstates which have $\omega$'s the closest to this value. } \tabularnewline \hline 
\texttt{residue\_conv\_thr } & $10E-4$ &
  {\footnotesize Threshhold for the convergence. When the square of the residue is smaller than this value, the convergence is achieved. } \tabularnewline \hline 
\texttt{start } & 0 &
  {\footnotesize The lower limit of the energy (\SB{Ry}) scale for the spectrum calculation. } \tabularnewline \hline 
\texttt{finish } & 1 &
  {\footnotesize The upper limit of the energy (\SB{Ry}) scale for the spectrum calculation. } \tabularnewline \hline 
\texttt{broadening } & 0.005 &
  {\footnotesize Value (\SB{Ry}) to broaden the spectrum.} \tabularnewline \hline 
\texttt{step } & 0.001 &
  {\footnotesize Energy step (\SB{Ry}) for the spectrum calculation. } \tabularnewline \hline 

\texttt{ltammd }
  & \emph{.false.}  & {\footnotesize When set to \emph{.true.} the
    Tamm-Dancoff approximation is used in constructing the
    Liouvillian.}\tabularnewline \hline 
\texttt{no\_hxc } &
  \emph{.false.}  & {\footnotesize When set to \emph{.true.} the
    change in the internal field (Hartree and exchange-correlation) is
    ignored in the calculation, resulting in an independent electron
    approximation.}\tabularnewline \hline 
{\bf \texttt{ecutfock }} &
  \texttt{ecutwfc} & {\footnotesize The charge density
    cut-off of the reduced grid used to calculate the EXX-like
    operators.}\tabularnewline \hline 
  {{\texttt{d0psi\_rs}}} & {{\footnotesize .false.}} & 
  {{\footnotesize When set to .true., the matrix elements of dipole
          operator are calculated in real space.  Can be used only for molecules 
          or clusters, NOT for periodic systems. Molecule must be placed in the 
          centre of the simulation cell. It is recommended to set this value
          only for hybrid-functional calculations.}} 
\tabularnewline \hline
  \end{tabular} \tabularnewline 
 \cline{1-1} 
\end{tabular} \tabularnewline
\caption{Input variables for \texttt{turbo\_davidson.x} }
\label{TableINPdavidson.x}
\end{table}

\newpage
\section{Sample inputs}
\begin{verbatim}
&control
    calculation='scf'
    restart_mode='from_scratch',
    pseudo_dir = './pseudo',
    outdir='./out'
    prefix='CO'
/
&system
    ibrav= 1,
    celldm(1)=20,
    nat=  2,
    ntyp= 2,
    ecutwfc = 40
    ecutfock=40
    input_dft='PBE0'
    nosym=.true.
    x_gamma_extrapolation=.false.
    exxdiv_treatment='vcut_spherical'
/
&electrons
    adaptive_thr=.true.
    conv_thr=1d-10
/
ATOMIC_SPECIES
 C 1.0 C.pbe-mt_gipaw.UPF
 O 1.0 O.pbe-mt.UPF
ATOMIC_POSITIONS {angstrom}
  C	5.000	5.000	4.436
  O	5.000	5.000	5.564
K_POINTS {gamma}
\end{verbatim}
{\footnotesize Input sample for {\texttt{pw.x}}
  calculation of a CO molecule using the PBE0 functional,
  as documented in the PWscf input guide in the \QE distribution.}
\newpage
\begin{verbatim}
&lr_input
    prefix="CO",
    outdir='./out',
    restart_step=500,
/
&lr_control
    itermax=1000,
    ipol=4,
    ecutfock=40d0 !This value has to be the same as in the scf input
    pseudo_hermitian=.true.   
    d0psi_rs = .true.
/
\end{verbatim}
{\footnotesize Input sample for {\texttt{turbo\_lanczos.x}}}

\begin{verbatim}
&lr_input
    prefix="CO",
    outdir='./out',
/
&lr_dav
    ecutfock=40d0,
    num_eign=20,
    num_init=40,
    num_basis_max=200,
    residue_conv_thr=1.0E-4,
    start=0.0,  
    finish=1.0,
    step=0.001,
    broadening=0.005,
    reference=0.0
    d0psi_rs = .true.
/
\end{verbatim}
{\footnotesize Input sample for {\texttt{turbo\_davidson.x}}}

\newpage


\end{document}